\documentclass[12pt]{article}
\usepackage{graphics}
\usepackage{amssymb}
\usepackage{psfrag}

\newcommand{\rf}[1]{(\ref{#1})}
\newcommand{\beq}{\begin{equation}}
\newcommand{\eeq}{\end{equation}}
\newcommand{\bea}{\begin{eqnarray}}
\newcommand{\eea}{\end{eqnarray}}

\newcommand{\e}{\mbox{e}}

\newcommand{\del}{\delta}
\newcommand{\Del}{\Delta}
\newcommand{\sg}{\sigma}

\newcommand{\prt}{\partial}
\newcommand{\mi}{\!-\!}
\newcommand{\equ}{\!=\!}
\newcommand{\pl}{\!+\!}

\newcommand{\cD}{{\cal D}}

\begin{document}

{\normalsize \hfill SPIN-05/05}\\
\vspace{-1.5cm}
{\normalsize \hfill ITP-UU-05/07}\\
${}$\\

\begin{center}
\vspace{48pt}
{ \Large \bf  Spectral Dimension of the Universe}

\vspace{40pt}

{\sl J. Ambj\o rn}$\,^{a,c}$,
{\sl J. Jurkiewicz}$\,^{b}$
and {\sl R. Loll}$\,^{c}$

\vspace{24pt}
{\footnotesize

$^a$~The Niels Bohr Institute, Copenhagen University\\
Blegdamsvej 17, DK-2100 Copenhagen \O , Denmark.\\
{ email: ambjorn@nbi.dk}\\

\vspace{10pt}

$^b$~Mark Kac Complex Systems Research Centre,\\
Marian Smoluchowski Institute of Physics, Jagellonian University,\\
Reymonta 4, PL 30-059 Krakow, Poland.\\
{email: jurkiewicz@th.if.uj.edu.pl}\\

\vspace{10pt}

$^c$~Institute for Theoretical Physics, Utrecht University, \\
Leuvenlaan 4, NL-3584 CE Utrecht, The Netherlands.\\
{email:  j.ambjorn@phys.uu.nl, r.loll@phys.uu.nl}\\

\vspace{10pt}

}
\vspace{48pt}

\end{center}

\begin{center}
{\bf Abstract}
\end{center}

\noindent
We measure the spectral dimension of
universes emerging from nonperturbative
quantum gravity, defined through state sums of causal
triangulated geometries. While four-dimensional
on large scales, the quantum universe appears two-di\-men\-sional at short
distances. We conclude that quantum gravity may be 
``self-renormalizing" at the Planck scale, by virtue of a
mechanism of dynamical dimen\-sional reduction. 

\vspace{12pt}
\noindent


\newpage

\subsection*{Quantum gravity as an ultraviolet regulator?}\label{intro}

A shared hope of researchers in otherwise disparate approaches to quantum
gravity is that the microstructure of space and time may provide a
physical regulator for the ultraviolet infinities encountered in perturbative
quantum field theory. The outstanding challenge is to construct {\it a}
consistent quantum description of this highly nonperturbative gravitational 
regime that stands a chance of being physically correct. 

Slow progress
in the quest for quantum gravity has not hindered speculation on what
kind of mechanism may be responsible for resolving the
short-distance singularities. A
recurrent idea is the existence of a minimal length scale,
often in terms of a characteristic Planck-scale unit of length in scenarios 
where the spacetime at short distances is fundamentally discrete.
An example is that of so-called loop quantum gravity,
where the discrete spectra of geometric operators measuring areas
and volumes on a kinematical Hilbert space are often taken as
evidence for fundamental discreteness in nature
\cite{loops1,loops2}.\footnote{Although this may be an attractive scenario, 
it is quite possible
that the discreteness (and associated finiteness) results 
of loop quantum gravity
will not survive the transition to the physical sector of the theory, which 
also solves the Hamiltonian constraint. One may also question the
genericity of the ``emergence" of such discreteness, which is 
a consequence \cite{hanno} of the insistence that one-dimensional,
unsmeared Wilson-line variables be represented by finite operators 
in the quantum theory.} Other quantization programs for gravity, such as
the ambitious causal set approach \cite{rafael}, postulate fundamental 
discreteness at the outset. 

The alternative we will advance here is based on new results
from an analysis of the properties of quantum universes generated
in the nonperturbative and background-independent CDT 
(causal dynamical triangulations) approach to quantum gravity. 
As shown in \cite{ajl-prl,semi}, they have a number of appealing
macroscopic properties: firstly, their scaling behaviour as function of the
spacetime volume is that of genuine
isotropic and homogeneous four-dimensional worlds. Secondly,
after integrating out all dynamical variables but the scale factor
$a(\tau)$ {\it in the full quantum theory}, the
correlation function between scale factors at different
(proper) times $\tau$ is described by the simplest
minisuperspace model used in quantum cosmology. 

We have recently begun an analysis of the {\it microscopic} properties
of these quantum spacetimes. As in previous work, 
their geometry can be probed in a rather direct manner
through Monte Carlo simulations and measurements. 
At small scales, it exhibits neither 
fundamental discreteness nor indication of a minimal length scale. 
Instead, we have found evidence of a fractal structure (see
\cite{new}, which also contains a detailed technical account of
the numerical set-up). What we report on in this letter is a most
remarkable finding concerning the universes' {\it spectral
dimension}, a diffeomorphism-invariant quantity obtained
from studying diffusion on the quantum ensemble of geometries.
On large scales and within measuring accuracy, it is equal to four, 
in agreement with earlier measurements of the
large-scale dimensionality based on the scale factor.
Surprisingly, the spectral dimension turns out to be scale-dependent
and decreases smoothly from four to a value of around two as the quantum
spacetime is probed at ever smaller distances. This suggests a
picture of physics at the Planck scale which is
radically different from frequently invoked scenarios of fundamental
discreteness: 
through the dynamical generation of a scale-dependent dimensionality,
nonperturbative quantum gravity provides an effective ultraviolet
cut-off through {\it dynamical dimensional reduction}.

\subsection*{The spectral dimension}

A diffusion process on a $d$-dimensional Euclidean geometry with
a fixed, smooth metric $g_{ab}(\xi)$ is governed by the diffusion
equation
\beq\label{ja2}
\frac{\prt}{\prt \sg} \, K_g(\xi,\xi_0;\sg) = \Del_g K_g (\xi,\xi_0;\sg),
\eeq
where $\sg$ is a fictitious diffusion time, $\Del_g$ the Laplace
operator corresponding to $g_{ab}(\xi)$ and $K_g(\xi,\xi_0;\sg)$
the probability density of diffusion from $\xi$ to $\xi_0$ in
diffusion time $\sg$. 
We will consider processes which are initially peaked at some point $\xi_0$,
\beq\label{ja3}
K_g(\xi,\xi_0;\sg=0) = \frac{\del^{d}(\xi-\xi_0)}{\sqrt{\det g(\xi)}}.
\eeq
A quantity that is easier to measure than $K_g$ in numerical
simulations is the average {\it return probability}
\beq\label{ja5}
P_{g}(\sg) := \frac{1}{V} \int d^d\xi \sqrt{\det g(\xi)} \; K_g(\xi,\xi;\sg),
\eeq
where $V\equ\int d^d\xi \sqrt{\det g(\xi)}$ is the spacetime volume.

For an infinite flat space,
the solution to eq.\ \rf{ja2} is simply given by
\beq\label{ja4}
K_g(\xi,\xi_0;\sg) = \frac{\e^{-d_g^2(\xi,\xi_0)/4\sg}}{(4\pi \sg)^{d/2}},
\qquad g_{ab}(\xi)\equ \delta_{ab},
\eeq
where $d_g(\xi,\xi_0)$ denotes the geodesic distance
between $\xi$ and $\xi_0$. It follows that $\sqrt{\sg}$ is an
effective measure of the linear spread of the Gaussian at
diffusion time $\sg$.
Because of $P_g(\sg)= 1/\sg^{d/2}$ in the flat case, we
can extract the dimension $d$ of the manifold by taking the logarithmic
derivative, 
\beq\label{ja5a}
-2\frac{d \log P_g(\sg)}{d\log \sg} = d,
\eeq
independent of $\sg$. 

For curved spaces and/or finite volume $V$ one can still use
eq.\ \rf{ja5a} to extract the dimension, but there will be
corrections for sufficiently large $\sg$.
For finite volume in particular, $P_g(\sg)$ goes to one for
$\sg \gg V^{2/d}$ since the zero mode of the Laplacian $-\Del_g$
will dominate the diffusion in this region. 
For a given diffusion time $\sg$ the behaviour of $P_g(\sg)$ is determined by
eigenvalues $\lambda_n$ of $-\Del_g$ with $\lambda_n \le 1/\sg$,
and the contribution from higher eigenvalues is exponentially suppressed.
Like in the flat case, large $\sg$ is related to diffusion which probes
spacetime at large scales, whereas small $\sg$ probes short distances.

We will use the return probability to determine an effective dimensionality 
of quantum spacetime, which in our nonperturbative path integral
formulation amounts to studying diffusion on an entire
{\it ensemble} of curved, Euclidean(ized) geometries.
Since the return probability $P_g(\sg)$ in \rf{ja5}
is invariant under reparametri\-zations, one can
define the quantum average $P_V(\sg)$ of $P_g(\sg)$ over all equivalence
classes $[g_{ab}] $ of metrics
with a given spacetime volume $V$ by
\beq\label{ja7}
P_V(\sg) =
\frac{1}{Z_V} \int\!\! \cD [g_{ab}] \; e^{-{S}_E(g_{ab})}
\del (\int d^4\xi \sqrt{\det g}-V) \, P_g(\sg),
\eeq
where $S_E(g_{ab})$ is the (Euclidean) Einstein-Hilbert action
of $g_{ab}(\xi)$ and $Z_V$ the partition function
(path integral) for geometries with fixed volume $V$.

It is straightforward to generalize the diffusion process and its
associated probability density to the piecewise linear geometries
that appear in the path integral of CDT \cite{new}. In analogy
with the ordinary path integral for a particle, one would expect
that in the continuum limit a typical geometry in the quantum
ensemble is still continuous, but nowhere differentiable;
for example, it could be fractal. In fact, diffusion on
fractal structures is well studied in statistical physics \cite{bah}, and
there the return probability takes the form
\beq\label{ja6}
P_N(\sg)= \sg^{-D_S/2} \; F\Big(\frac{\sg}{N^{2/D_S}}\Big),
\eeq
where $N$ is the ``volume'' associated with the fractal
structure and $D_S$ is the so-called {\it spectral dimension},
which is not necessarily an integer.
An example of fractal structures are branched polymers,
which generically have a spectral dimension $D_S\equ 4/3$
\cite{thordur-john,anrw}.

For the nonperturbative gravitational path integral defined by CDT,
we expect the same functional form \rf{ja6}, where $N$ now stands
for the discrete four-volume, given in terms of the number of 
four-simplices of a triangulation.
As above in \rf{ja5a}, we can now extract 
the value of the fractal dimension $D_S$ by
measuring the logarithmic derivative,
\beq\label{ja1}
D_S(\sg) = -2 \;\frac{d\log P_N(\sg)}{d\log \sg},
\eeq
as long
as the diffusion time $\sg$ is not much larger than $N^{2/D_S}$.
From previous numerical simulations of 2d quantum gravity 
in terms of dynamical triangulations \cite{ajw,2dspectral,aa}
we already know that eq.\ \rf{ja1} is not reliable for arbitrary
diffusion times. The observed behaviour of $D_S(\sg)$ for a given
triangulation will typically exhibit irregularities
for the smallest $\sg$, caused by the lattice discretization\footnote{Often the
behaviour of $P_N(\sg)$ for odd and even numbers of diffusion steps $\sg$ will
be quite different for small $\sg$ and merge only
for $\sg \approx 20-30$. -- The origin of this asymmetry is illustrated 
by the (extreme) example of diffusion on a one-dimensional piecewise straight 
space, where the return probability
simply vanishes for any odd number $\sg$ of steps.\label{foot1}}, and then enter 
a long and stable regime where the spectral dimension is independent of $\sg$,
before finite-size effects start to dominate and $D_S(\sg)$ goes
to zero. 

\subsection*{Measuring the spectral dimension}

In the CDT approach, quantum gravity is defined as the continuum 
limit of a regularized version of the nonperturbative gravitational path 
integral \cite{ajl1, ajl4d}. 
The set of spacetime geometries to be summed over is represented
by a class of causal four-dimensional piecewise flat manifolds 
(``triangulations"). Every member $T$ of the ensemble of simplicial
spacetimes can be wick-rotated to a unique Euclidean piecewise
flat geometry, whereupon the path integral assumes the form of a
partition function
\begin{equation}
Z=\sum_T \frac{1}{C_T} {\rm e}^{-S_E(T)},
\label{parti}
\end{equation}
where $C_T$ is a combinatorial symmetry factor and $S_E(T)$
the Euclidean Einstein-Regge action of the triangulation $T$
\cite{ajl4d}. All geometries share a global, discrete version of
proper time. In the continuum limit, the CDT time $\tau$ becomes proportional to the
cosmological proper time of a conventional minisuperspace model
\cite{semi}.\footnote{The proper time $\tau$ of CDT --
although invariantly defined from a geometric point of view --
does not necessarily coincide with a physical time appearing 
explicitly in the formulation of physical observables, for example, two-point functions.} 

We extract information about the continuum limit of the theory
by Monte Carlo simulations and a finite-size scaling analysis of
\rf{parti}. Further details about the updating moves for the geometry
and the numerical set-up can be found in \cite{ajl4d} and \cite{new}
respectively.
For computer-technical reasons we keep the total spacetime
volume $N$ approximately fixed during simulations. All the results
obtained can be related by an inverse Laplace transform to
those for the geometric ensemble with the volume constraint absent.

Our measurements of the spectral dimension were performed in
the phase of the statistical model \rf{parti} which generates
a quantum geometry extended in both space and time,
with large-scale four-dimensional scaling properties \cite{ajl-prl,new}. 
As discussed in \cite{semi}, the quantum universe we generate
is a ``bounce" in the Euclidean sector of the theory. It has a
characteristic shape when we plot its three-volume (equivalently, its
scale factor $a(\tau)$) as a function of proper time $\tau$.
The universe starts out with a minimal three-volume, increases
to a maximum, and then decreases in a symmetric fashion back to 
a minimal value. As shown in \cite{semi}, the underlying dynamics
of the scale factor $a(\tau)$ solves the equation of motion 
of a simple minisuperspace action
for $a(\tau)$, and is that of a bounce, with total Euclidean spacetime
volume determined by the volume at which the simulation is
performed. 
It is on an ensemble of geometries of this type that we 
made our measurements.
Since we are interested in the bulk properties of the diffusion,
we always started the process from a simplex in the constant-time
slice where $a(\tau)$ is maximal.
The simulations to determine the universe's spacetime spectral dimension 
were performed for geometries of proper-time extension $t\equ 80$
and a discrete volume of up to approximately $N\equ 181.000$ four-simplices,
and the diffusion was followed for up to $\sg_{\rm max}\equ 400$ time steps.

\begin{figure}[t]
\psfrag{X}{{$\;\;\;\sg$}}
\psfrag{Y}{{ $\!\!\! D_S$}}
\centerline{\scalebox{1.2}{\rotatebox{0}{\includegraphics{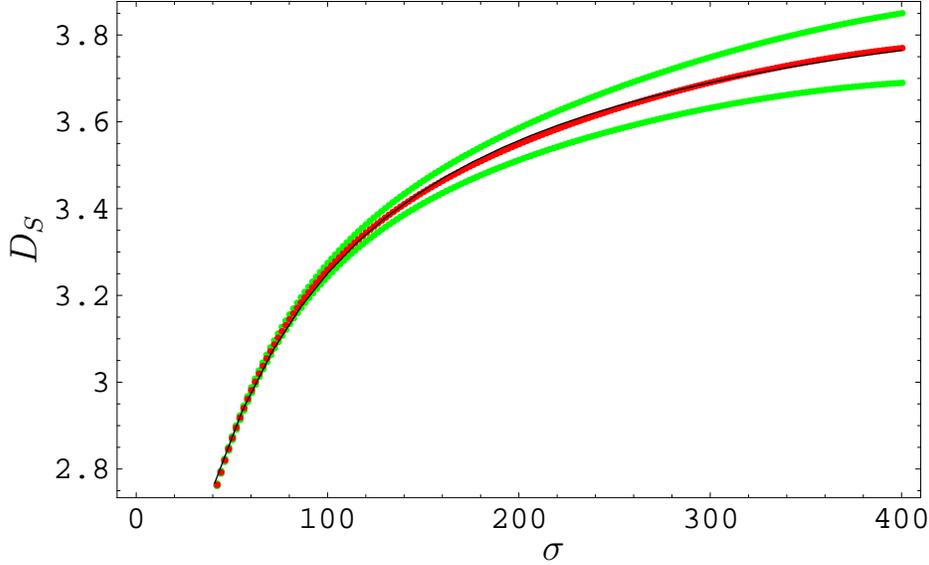}}}~~~~
~~~}
\caption[phased]{{\small The data points along the central curve show the 
spectral dimension $D_S(\sg)$ of the universe as function
of the diffusion time $\sg$. Superimposed is a best fit, the continuous curve
$D_S(\sg) = 4.02\mi 119/(54\pl\sg)$. The two outer curves quantify the error bars, which 
increase linearly with $\sg$, due to \rf{ja1}. 
(Measurements taken for a quantum universe with 181.000 
four-simplices.) }}
\label{d4s2.2b4}
\end{figure}

Fig.\ \ref{d4s2.2b4} summarizes our measurements of the spectral
dimension $D_S(\sg)$ at the maximal spacetime volume, 
extracted as the logarithmic derivative \rf{ja1} from a
discrete implementation of the diffusion process. The (envelopes of the)
error bars represent the errors coming from averaging
over 400 different measurements of diffusion processes, performed for
independent starting points and statistically independent configurations
$T$ generated by the Monte Carlo simulation.
As observed previously in other systems of random geometry,
we have found a different behaviour of $D_S(\sg)$ for odd and even
(discrete) diffusion times $\sg$ for small $\sg$ (cf. footnote \ref{foot1}). 
In order to eliminate this short-distance lattice artifact, we have only
included the region $\sg \geq 40$ for which the odd and even curves 
coincide, both in Fig.\ \ref{d4s2.2b4} and in determining the
spectral dimension.

The data points along the central curve 
in Fig.\ \ref{d4s2.2b4} represent our best approximation 
to $D_S(\sg)$ in the limit of infinite spacetime volume. Their monotonic increase 
as function of $\sg$ indicates that we have not yet reached the region
where finite-volume effects dominate (in the form of the constant mode of
the Laplacian). The remarkable feature of the behaviour of the
spectral dimension illustrated in Fig.\ \ref{d4s2.2b4} is that it
is qualitatively different from what has been observed in similar
systems up to now, be it in two-dimensional Euclidean quantum
gravity with or without matter \cite{ajw} or for the spatial hypermanifolds of 
our present CDT set-up in four dimensions \cite{new}. In these cases,
immediately following the region of even-odd asymmetry for small $\sg$,
$D_S(\sg)$ stabilizes in a horizontal line, indicating the presence
of a single value $D_S^{(0)}$ characterizing the spectral dimension of the
system, independent of the scale at
which the diffusion process probes the geometry.

Apparently, this is not the case for the spectral spacetime dimension in
quantum gravity defined by CDT. The measurements shown in Fig.\ \ref{d4s2.2b4} 
indicate that $D_S(\sg)$ changes with the scale probed.
In order to quantify this scale dependence we have attempted
a variety of fits in the available data range $\sg \in [40,400]$.
Among curves with three free parameters of the form {\it const.}$+$asymptotic 
form\footnote{The two alternatives considered were $a\mi b{\rm e}^{-c\sg}$
and $a\mi b/\sg^c$.}, a fit of the form
\beq\label{ja8}
-2\,\frac{d \log P(\sg)}{d\log \sg} = a -\frac{b}{\sg+c}
\eeq
agrees best with the data. In Fig.\ \ref{d4s2.2b4}, the curve
\beq\label{ja9}
D_S(\sg) = 4.02-\frac{119}{54+\sg}
\eeq
has been superimposed on the data, where the three constants were 
determined from the entire data range $\sg \in [40,400]$. 
Although both $b$ and $c$ individually are slightly altered when one
varies the range of $\sg$, their ratio $b/c$ as well as the constant
$a$ remain fairly stable. Integrating relation \rf{ja8}, we have
\beq\label{ja8a}
P(\sg) \sim \frac{1}{\sg^{a/2} (1+c/\sg)^{b/2c}},
\eeq
implying a behaviour
\beq\label{ja8b}
P(\sg) \sim \left\{
\begin{array}{cl}
\displaystyle{{\sg^{-a/2}}} &~~\mbox{for large $\sg$,}\\
~&~\\
\displaystyle{{\sg^{-(a-b/c)/2}}} &~~ \mbox{for small $\sg$}.
\end{array}
\right.
\eeq
Our interpretation of eqs.\ \rf{ja8a} and \rf{ja8b} is that
{\it the quantum geometry generated by CDT does not have a 
self-similar structure at all distances, but instead has
a scale-dependent 
spectral dimension which increases continuously from
$a\mi b/c$ to $a$ with increasing distance.}

Taking into account the variation of $a$ in eq.\ \rf{ja8} when using
various cuts $[\sg_{\rm min}, \sg_{\rm max}]$ for the range of $\sg$, 
as well as different weightings of the errors, 
we obtain the asymptotic value
\beq\label{ja10}
D_S(\sg\equ \infty) = 4.02 \pm 0.1,
\eeq
which means that the spectral dimension extracted from
the large-$\sg$ behaviour (which probes the long-distance structure
of spacetime) is compatible with four. On the other hand, the 
``short-distance spectral dimension'',
obtained by extrapolating eq.\ \rf{ja8a} to $\sg \to 0$ is given by
\beq\label{ja11}
D_S(\sg\equ 0)= 1.80 \pm 0.25,
\eeq
and thus is compatible with the integer value two. 

\subsection*{Discussion}

The continuous change of spectral dimension described in this letter 
constitutes to our knowledge
the first dynamical derivation of a scale-dependent dimension
in full quantum gravity.\footnote{In the so-called exact
renormalization group approach to Euclidean quantum gravity,
a similar reduction has been observed recently in an
Einstein-Hilbert truncation \cite{lauscher}.} It is natural to conjecture it will
provide an effective short-distance cut-off by which the
nonperturbative formulation of quantum gravity employed here,
causal dynamical triangulations, evades the ultraviolet infinities
of perturbative quantum gravity. Contrary to current folklore (see 
\cite{garay} for a review), 
this is done without appealing to short-scale discreteness
or abandoning geometric concepts altogether.

Translating our lattice
results to a continuum notation requires a ``dimensional
transmutation" to dimensionful quantities, 
in accordance with the renormalization of the lattice theory. 
Because of the perturbative nonrenormalizability of gravity, this is expected
to be quite subtle. CDT provides a concrete framework for addressing
this issue and we will return to it elsewhere.
However, since $\sg$ from \rf{ja2} can be assigned the
length dimension two, and since we expect the short-distance behaviour
of the theory to be governed by the continuum gravitational coupling $G_N$,
it {\it is} tempting to write the continuum version of \rf{ja8} as
\beq\label{jb5}
P_V(\sg)  \sim \frac{1}{\sg^2}\;  \frac{1}{1+{\it const.}\cdot G_N/\sg},
\eeq
where {\it const.} is a constant of order one. The relation \rf{jb5} describes a
universe whose spectral dimension is four on scales large compared to the
Planck scale. Below this scale, the quantum-gravitational excitations of
geometry lead to a nonperturbative dynamical dimensional reduction to 
two, a dimensionality where gravity is known to be 
renormalizable.

\end{document}